\documentclass{article}

\usepackage{arxiv}

\usepackage[utf8]{inputenc} 
\usepackage[T1]{fontenc}    
\usepackage{hyperref}       
\usepackage{url}            
\usepackage{booktabs}       
\usepackage{amsfonts}       
\usepackage{nicefrac}       
\usepackage{microtype}      
\usepackage{lipsum}		
\usepackage{graphicx}
\usepackage{natbib}
\usepackage{doi}
\usepackage{xcolor}
\usepackage{soul}
\usepackage{svg}
\usepackage{comment}
\usepackage{placeins}

\hypersetup{
    colorlinks=true,
    linkcolor=blue,
    filecolor=blue,
    urlcolor=blue,
    citecolor=blue
}

\title{\textbf{GeoSim.AI}: AI assistants for numerical simulations in geomechanics}

\date{} 					

\author{ \href{https://orcid.org/0000-0002-7803-1283}{\includegraphics[scale=0.06]{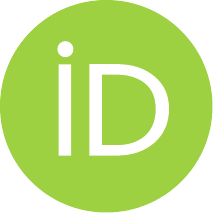}\hspace{1mm}Yared W. Bekele}\thanks{Corresponding author: yared.bekele@sintef.no.} \\
	Rock and Soil Mechanics Group\\
	Infrastructure Department\\
	SINTEF Community, Trondheim \\
	\texttt{yared.bekele@sintef.no} \\
}

\renewcommand{\headeright}{}
\renewcommand{\undertitle}{}
\renewcommand{\shorttitle}{\textit{GeoSim.AI}: AI assistants for numerical simulations in geomechanics}

\hypersetup{
pdftitle={GeoSim.AI: AI assistants for numerical simulations in geomechanics},
pdfsubject={},
pdfauthor={Yared W. Bekele},
pdfkeywords={AI, autonomous simulation, computational geomechanics, LLMs},
}

\begin{document}
\maketitle

\begin{abstract}
	The ability to accomplish tasks via natural language instructions is one of the most efficient forms of interaction between humans and technology. This efficiency has been translated into practical applications with generative AI tools now allowing users to get things done through natural language queries. The emergence of advanced Large Language Models (LLMs) marks a pivotal shift in this direction. With ongoing advancements in the field of generative AI, integrating natural language commands into sophisticated technical fields in science and engineering is becoming increasingly feasible. This paper introduces GeoSim.AI - a suite of AI assistants for numerical simulations in geomechanics - thereby demonstrating the transformative potential of generative AI in geotechnical engineering. We investigate how AI assistants powered by LLMs can streamline the process of creating complex simulation inputs and interpreting results by translating natural language instructions or image inputs into precise technical commands and scripts. This approach aims to bridge the gap between human intent and the intricate requirements of numerical modeling tools, potentially revolutionizing how researchers and engineers interact with simulation software. We present demonstrations involving AI assistants for performing slope stability analyses in various software packages. The demonstrations highlight the potential of this technology to significantly enhance productivity and accessibility in computational geomechanics. GeoSim.AI is under active development, continuously expanding the suite of AI assistants for various numerical simulation problems in geotechnical engineering.  
\end{abstract}


\section{Introduction}
The ability to accomplish tasks via natural language instructions is one of the most efficient forms of interaction between humans and technology. In recent years, the field of artificial intelligence (AI) has made remarkable strides, particularly in the domain of generative AI. The emergence of Large Language Models (LLMs) has revolutionized how humans interact with technology, enabling more intuitive and efficient communication through natural language. These advancements have found applications in various sectors and are continuously finding new areas of application in technical domains. 

The integration of AI-powered natural language interfaces with specialized technical domains represents a new frontier in human-computer interaction. This convergence has the potential to improve the usability of complex tools, methodologies and knowledge, making them more accessible to a broader range of users and applications. As AI continues to advance, the possibility of integrating natural language commands into sophisticated technical fields in science and engineering is becoming increasingly feasible. While this is an emerging field of application, the potential is demonstrated for various fields by recent research works. \cite{zhang2024geogpt} demonstrated the application of generative AI for geospatial applications. \cite{wiegand2024using} used generative AI to build an assistant for setting up reservoir simulations. \cite{alexiadis2024text} discussed how LLMs can be used together with simulation software for physics-based modelling.

Another field that stands to benefit significantly from this technological advancement is computational geomechanics. This discipline plays a crucial role in understanding and predicting the behavior of geological materials such as soils and rocks under various environmental conditions. Computational geomechanics relies heavily on numerical simulations to model complex phenomena based on software tools that often require significant time and effort to master. This presents a significant barrier to entry for students, researchers and engineers, requiring extensive domain knowledge and sometimes programming expertise to set up, run, and interpret the results effectively. In this context, the integration of generative AI assistants could significantly reduce the learning curve associated with numerical simulation software and improve productivity. This would enable researchers and engineers to focus more on the physics of the problem rather than the intricacies of simulation setup and execution, which is often different in different software packages for the same physical problem.

This paper introduces GeoSim.AI, a suite of novel AI assistants that present an innovative approach to streamline and enhance numerical simulations in computational geomechanics. GeoSim.AI leverages LLM-powered AI to bridge the gap between human intent expressed in natural language and the precise technical requirements of geomechanical simulation software. As a tool, GeoSim.AI provides a user-friendly chat interface that accepts natural language and image inputs. As an approach, it employs generative AI technologies to translate these inputs into accurate simulation parameters and automate the generation of complex scripts to setup numerical simulations.

One of the main aims of this paper is to explore GeoSim.AI's potential to revolutionize the field of computational geomechanics. We seek to investigate the feasibility of using Large Language Models to interpret natural language instructions for geomechanical simulations, demonstrating how GeoSim.AI can effectively translate user queries into precise simulation inputs and analysis scripts. Through this exploration, we will evaluate both the potential benefits and challenges of integrating AI assistants into computational geomechanics workflows. Additionally, we aim to examine the broader implications of this technology for research productivity and accessibility in the field, considering how tools like GeoSim.AI could help users focus more on the practical aspect of the problem while the complex geomechanical modeling process is handled by an AI assistant. We also address cross-cutting issues such as whether this approach means adding another layer of blackbox on top of simulation software that may already be considered blackboxes for users without a detailed knowledge of how the software tools work in the background. 

By addressing these objectives, we aim to contribute to the ongoing dialogue about the role of AI in scientific computing and engineering, specifically focusing on its application in geomechanics. The successful implementation of such a system could not only enhance the efficiency of experienced researchers but also lower the entry barrier for newcomers to the field, potentially accelerating the pace of discovery and innovation in geomechanical research and applications.

In the following sections, we will provide a background on the methodology behind GeoSim.AI's development, present a demonstrative application of the tool, discuss our findings, and conclude with reflections on the future prospects of AI-assisted simulation workflows in geomechanics and beyond, as well as on some cross-cutting implications.

\section{Architecture of GeoSimAI}

\begin{figure}
    \centering
    \includegraphics[scale=0.23]{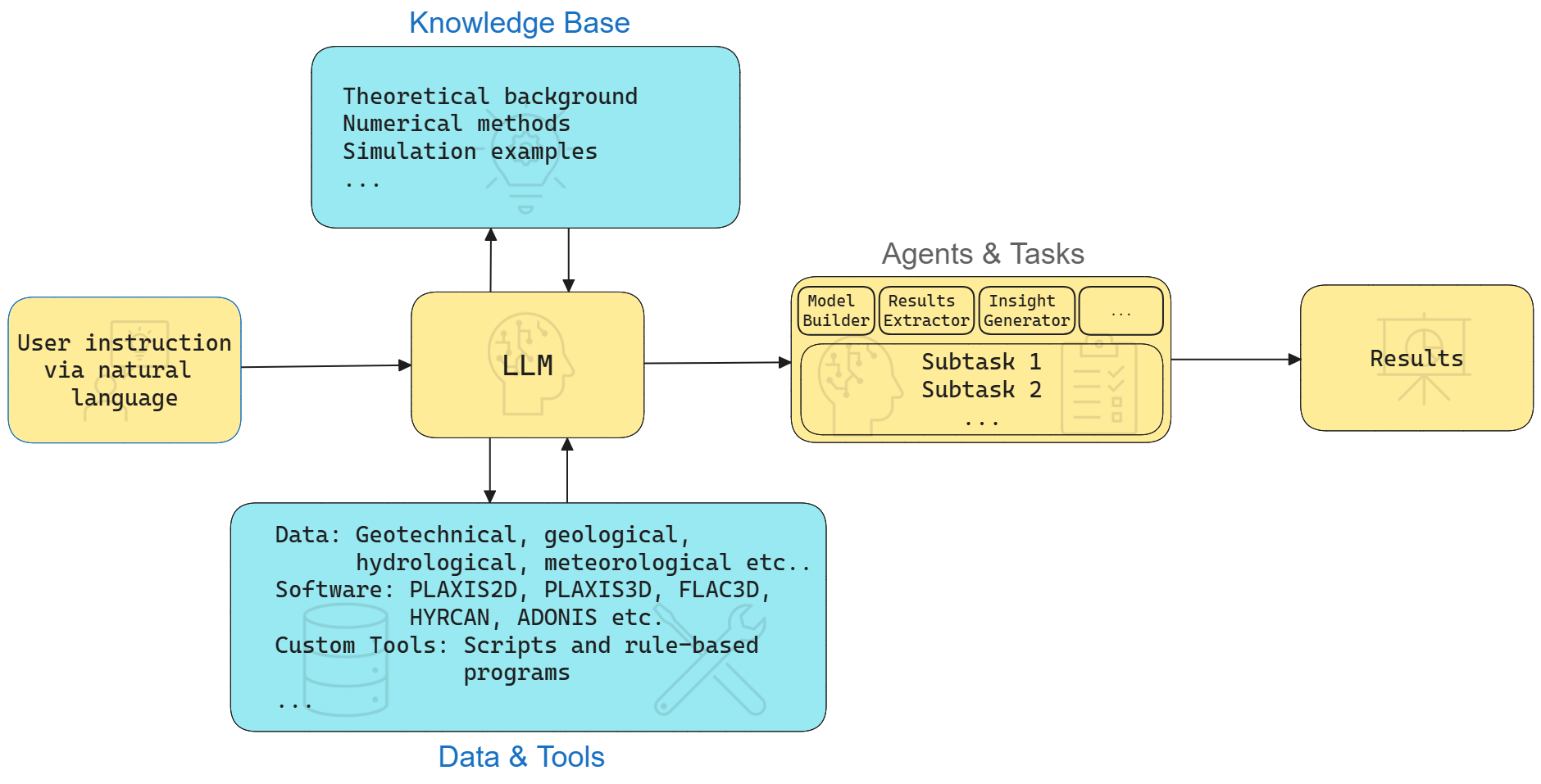}
    \caption{Architecture of GeoSim.AI.}
    \label{fig:GeoSimAI_architecture}
\end{figure}

The GeoSim.AI system architecture represents a novel integration of artificial intelligence with traditional geomechanical simulation workflows; see Figure \ref{fig:GeoSimAI_architecture}. Central to the system is a Large Language Model (LLM), which serves as the primary interface between user intent and technical execution. The LLM is complemented by two critical knowledge repositories: a comprehensive \emph{Knowledge Base} containing theoretical and practical information on geomechanics, and a \emph{Data \& Tools} base that houses relevant datasets and target software tools for simulations. At its core, the architecture is designed to bridge the gap between natural language instructions and complex numerical simulations in geomechanics. This section provides a detailed overview of the system's architecture.

\subsection{Natural Language Interface}
The Natural Language Interface of GeoSim.AI is implemented as a straightforward chat interface, providing users with a familiar and intuitive means of interaction. This interface serves as the primary point of entry for users to communicate their geomechanical simulation requirements to the system. This interface is designed to accept and interpret a wide range of geomechanical simulation requests expressed in natural language, including text, images and sketches, making the system accessible to users with varying levels of technical expertise. These inputs can range from simple queries to complex descriptions of geomechanical scenarios, including specifications of material properties, boundary conditions, loading scenarios, and desired outputs. The interface also provides users with the option to select their preferred simulation software from a list of supported tools.

\subsection{Large Language Model (LLM) Core}
The Large Language Model (LLM) serves as the central processing unit of GeoSim.AI, interpreting user requests and orchestrating the simulation workflow. This core component is designed with flexibility and adaptability in mind, allowing for the integration of different state-of-the-art LLMs as they become available. 

GeoSim.AI's architecture is developed to be LLM-agnostic, enabling the system to leverage the most suitable or advanced LLM for its purposes. This flexibility ensures that the system can evolve alongside rapid advancements in AI technology, maintaining its effectiveness and relevance over time.

A key challenge in applying general-purpose LLMs to specialized domains like geomechanics is the need for domain-specific knowledge. Out-of-the-box LLMs, while powerful, typically lack the depth of understanding required for complex geomechanical simulations. To address this, GeoSim.AI employs Retrieval Augmented Generation (RAG) to enhance the LLM's capabilities.

The RAG approach allows the LLM to dynamically access and incorporate relevant information from external sources during its reasoning and generation processes. In GeoSim.AI, this is achieved by providing context to the LLM via a comprehensive Knowledge Base containing theoretical foundations, best practices, and case studies in geomechanics, as well as a Data \& Tools Base that houses relevant datasets, software information, and customized rule-based scripts.

By leveraging RAG, the LLM can effectively combine its general language understanding and reasoning capabilities with specific, up-to-date geomechanical knowledge. This approach enables GeoSim.AI to provide accurate and contextually relevant responses to user queries, translate natural language instructions into technical simulation inputs, and guide the overall simulation process.

While the current implementation of GeoSim.AI relies on RAG, an alternative approach to enhance LLM capabilities is fine-tuning on specialized geomechanical knowledge datasets. This fine-tuning approach is a subject for future exploration and development, potentially offering different performance characteristics in geomechanical applications.

\subsection{Knowledge Base}
The Knowledge Base is a critical component of GeoSim.AI, serving as a comprehensive repository of geotechnical engineering knowledge and practical examples. This carefully curated collection of documents and resources forms the foundation upon which the Large Language Model (LLM) builds its understanding and decision-making processes in the context of geomechanical simulations.

At its core, the Knowledge Base encompasses the fundamentals of numerical modeling in geotechnical engineering. This includes detailed information on various numerical methods such as finite element analysis, finite difference methods, and discrete element methods. The theoretical background for these methods is thoroughly documented, providing the LLM with a solid grounding in the mathematical and physical principles underlying geomechanical simulations.

Beyond theory, the Knowledge Base also contains extensive documentation on best practices in numerical modeling. This practical knowledge covers aspects such as mesh generation, boundary condition selection, constitutive model choice, and result interpretation. By incorporating these best practices, GeoSim.AI can guide users towards creating more accurate and reliable simulations, even if they are not experts in numerical modeling techniques.

A key feature of the Knowledge Base is its collection of example numerical problems. These examples are tailored to showcase the command and scripting syntax of various popular geomechanics software packages, including ADONIS, HYRCAN, FLAC, and PLAXIS. By including a diverse range of examples, the Knowledge Base enables GeoSim.AI to provide software-specific guidance, translating user requirements into the appropriate syntax and structure for their chosen simulation tool.

To maximize the utility of this vast repository of information, the documents within the Knowledge Base are meticulously indexed and converted into vector representations. This process creates vector stores that the LLM can efficiently query based on the user's natural language requests. The vectorization allows for semantic search capabilities, enabling the LLM to retrieve not just keyword matches, but conceptually relevant information that may be expressed in different terms.

This indexing and retrieval system forms the backbone of GeoSim.AI's Retrieval Augmented Generation (RAG) approach. When a user presents a query or describes a simulation scenario, the LLM can swiftly access and incorporate the most relevant information from the Knowledge Base. This dynamic integration of curated knowledge with the LLM's inherent language understanding capabilities allows GeoSim.AI to provide responses that are both linguistically coherent and technically accurate in the domain of geotechnical engineering.

As the GeoSim.AI system is under active development, the knowledge base curation is an ongoing process which starts small with a focused application for slope stability problems.

\subsection{Data and Tools}
The Data and Tools repository is an essential component of GeoSim.AI, complementing the Knowledge Base by providing practical resources for simulation tasks. This repository is designed to enhance the system's ability to respond to user requests with realistic data and perform necessary calculations and analyses.

GeoSim.AI's Data Repository comprises a diverse collection of typical datasets used in various modelling scenarios within geotechnical engineering. These datasets span multiple domains, including geotechnical, geological, hydrological, and meteorological data. The inclusion of such varied data enables GeoSim.AI to respond to user requests in a sensible and context-appropriate manner.

One of the key advantages of this data repository is its ability to assist the system in making necessary assumptions when faced with limited data availability from the user. By leveraging these typical datasets, GeoSim.AI can fill in gaps in user-provided information with realistic values, ensuring that simulation inputs can be generated even with incomplete initial data. This feature is particularly valuable for preliminary analyses or when users are in the early stages of project planning.

Complementing the data repository is a comprehensive set of tools tailored for various purposes in the geomechanical simulation workflow. These tools include scripts and programs for performing preliminary calculations, extracting results from simulations, and generating plots and visualizations. Importantly, these tools are customized to work with different geomechanical software packages, ensuring compatibility with the user's chosen simulation environment.

A significant feature of the Tools Repository is its ability to generate insights from simulation results. Users can provide simulation output files, and GeoSim.AI can utilize appropriate tools to analyze these results, extract key information, and present findings in a user-friendly format. This capability extends the utility of GeoSim.AI beyond the simulation setup phase, making it a valuable asset throughout the entire modelling process.

Another important aspect of the Data and Tools Repository is its integration with the Large Language Model (LLM) through the LLM's Function Calling feature. This advanced capability of modern LLMs allows them to dynamically access and utilize specific tools or datasets as needed during their interaction with the user. For instance, when a user requests a particular analysis or visualization, the LLM can identify the appropriate tool within GeoSim.AI's repository, call the necessary function, and incorporate the results into its response.

The Function Calling feature of LLMs significantly enhances GeoSim.AI's capabilities by enabling it to perform concrete actions beyond model building command or script generation. It allows the system to leverage the LLM's ability to run data processing scripts or generate plots in real-time, based on the context of the user's request. This seamless integration of the LLM's language understanding and practical tool usage makes GeoSim.AI a powerful assistant capable of providing not just modelling guidance but also tangible, data-driven insights.

The combination of typical datasets, specialized tools, and the LLM's function calling capabilities enables GeoSim.AI to offer comprehensive support throughout the geomechanical simulation process, from initial data input to final result interpretation.

In the current version of GeoSim.AI, the resources in the Data \& Tools repository are limited and under continuous improvement. Connection of the LLM to the repository, for example to run simulation using a chose software or extract results from an already ran simulation, is under test and development.  

\section{Demonstration Examples}

\subsection{GeoSim.AI Interface and General Demonstration}

\begin{figure}
    \centering
    \includegraphics[scale=0.38]{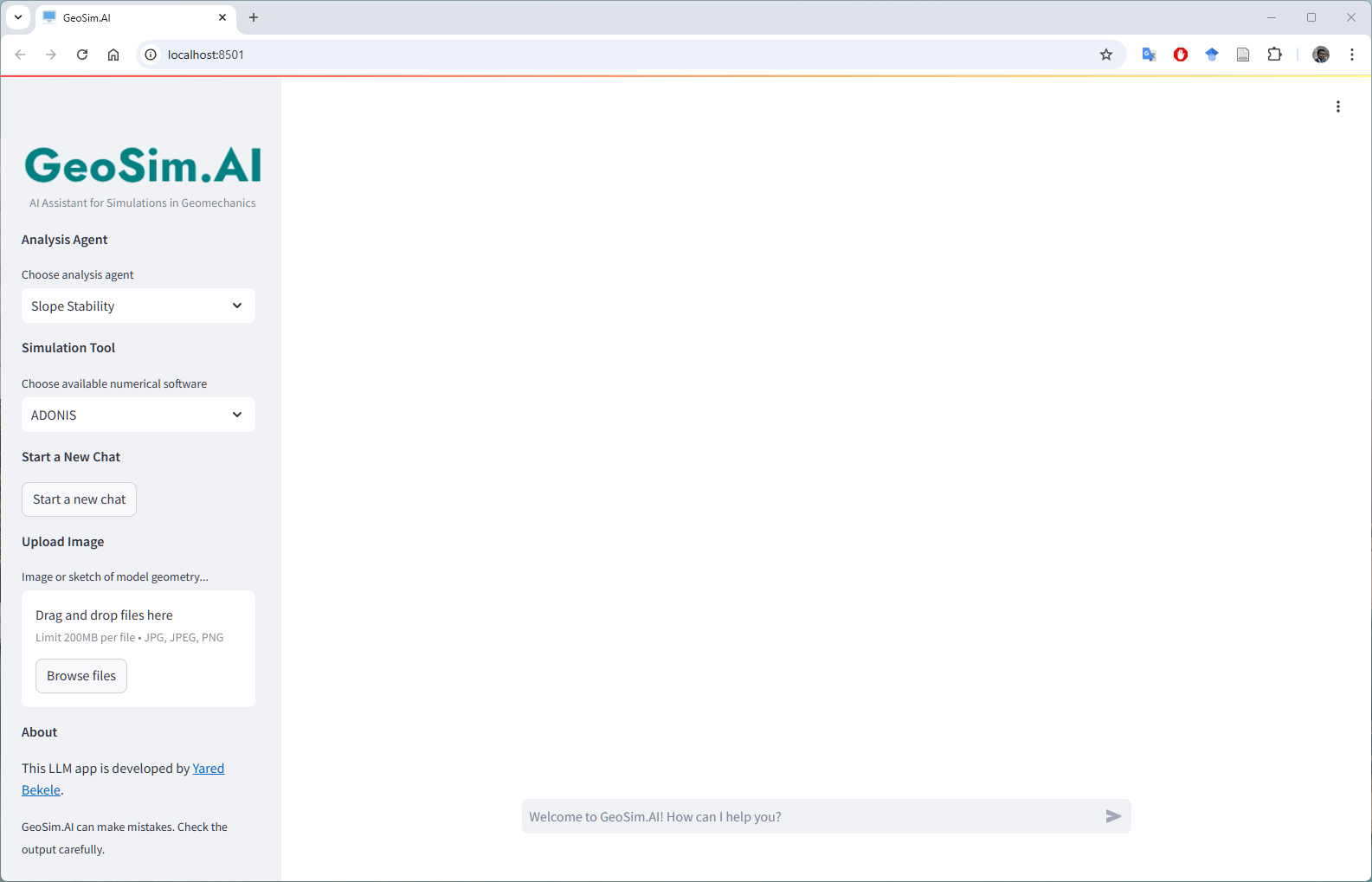}
    \caption{Chat Interface of GeoSim.AI.}
    \label{fig:geosimai_interface}
\end{figure}

The GeoSim.AI interface, shown in Figure~\ref{fig:geosimai_interface}, features a side panel that provides the selection of agents and tools. Users can select an analysis agent and a numerical simulation tool from dropdown menus, allowing flexibility in the type of simulation performed and the choice of a target software. Additionally, the interface includes an image upload option that supports multiple file formats, enabling the input of model geometries for analysis. A chat interface at the bottom allows users to interact with the AI assistant.

\begin{figure}
    \centering
    \includegraphics[scale=0.38]{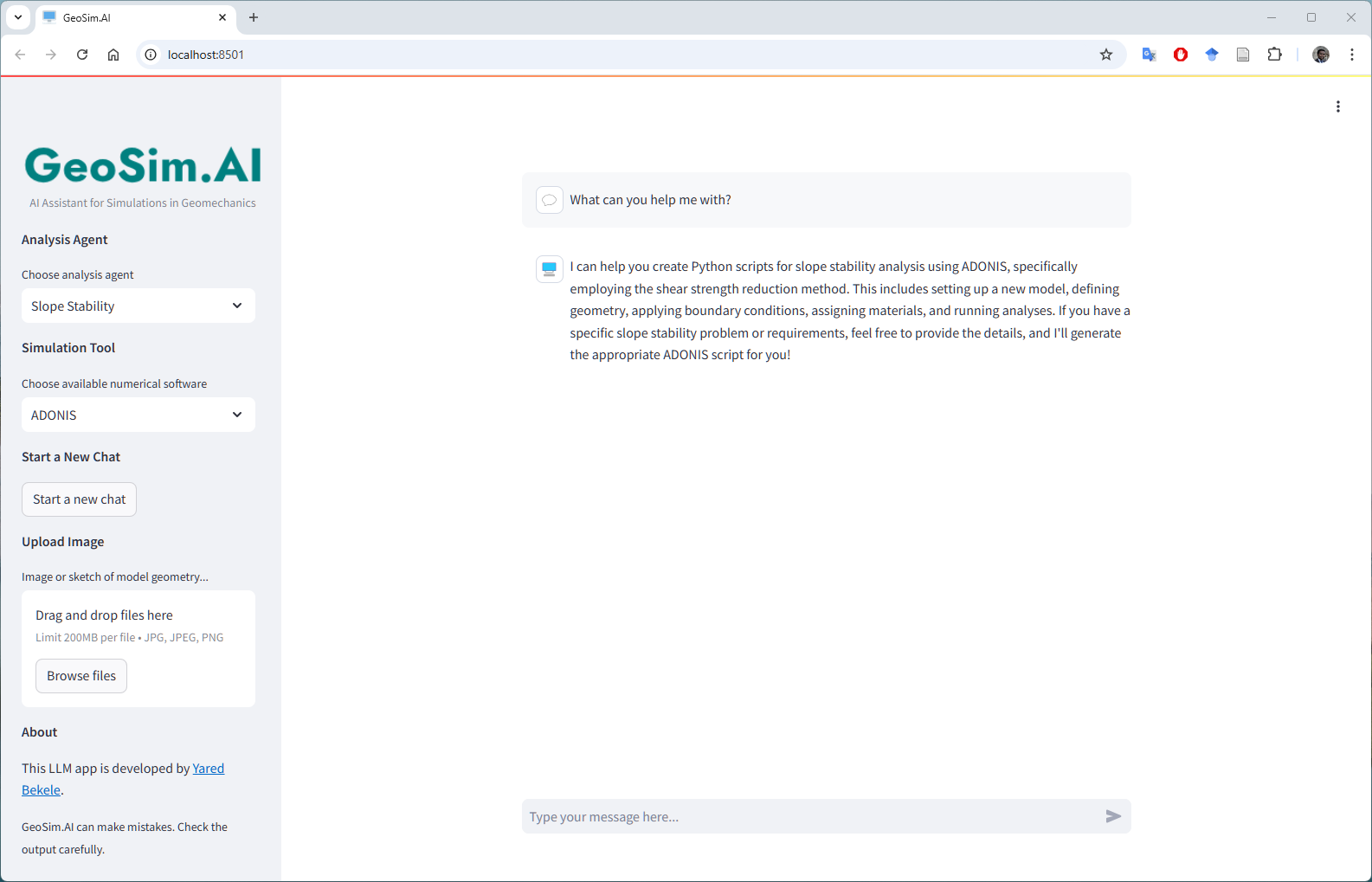}
    \caption{Example chat with GeoSim.AI - slope stability assistant for ADONIS.}
    \label{fig:general_demo_1}
\end{figure}

\begin{figure}
    \centering
    \includegraphics[scale=0.38]{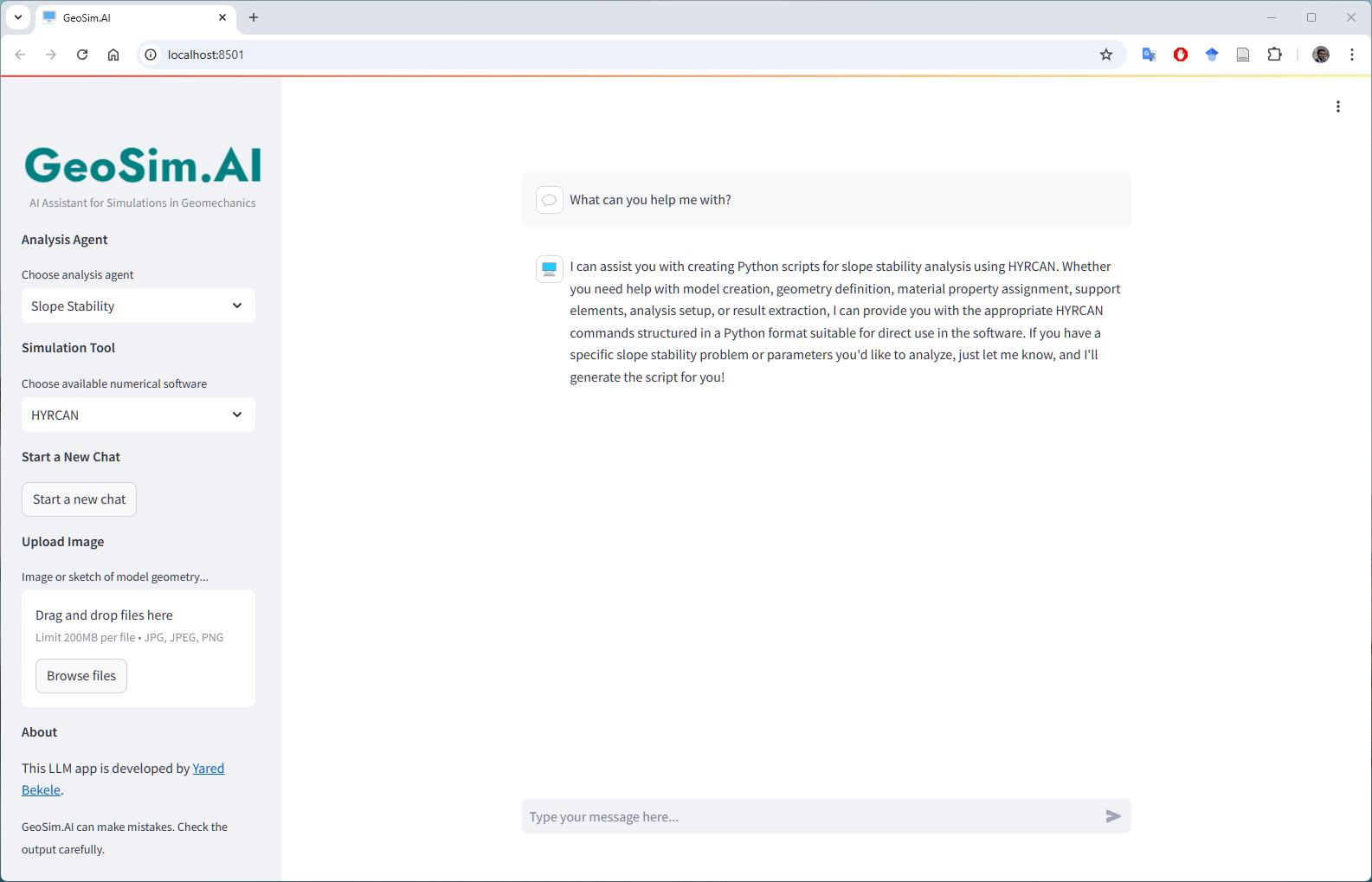}
    \caption{Example chat with GeoSim.AI - slope stability assistant for HYRCAN.}
    \label{fig:general_demo_2}
\end{figure}

Figure~\ref{fig:general_demo_1} shows an example chat where we ask GeoSim.AI the question "What can you help me with?" with the analysis agent set to Slope Stability and the target software is ADONIS. Similarly, Figure~\ref{fig:general_demo_2} shows an example chat with the slope stability assistant for HYRCAN. A video version of the demonstrations can be seen by following this link: \href{https://youtu.be/_LprVXHBT-I}{GeoSim.AI General Demo}.

\FloatBarrier

\subsection{AI Assistant for ADONIS}

\subsubsection*{Text Prompt}

\begin{figure}
    \centering
    \includegraphics[scale=0.32]{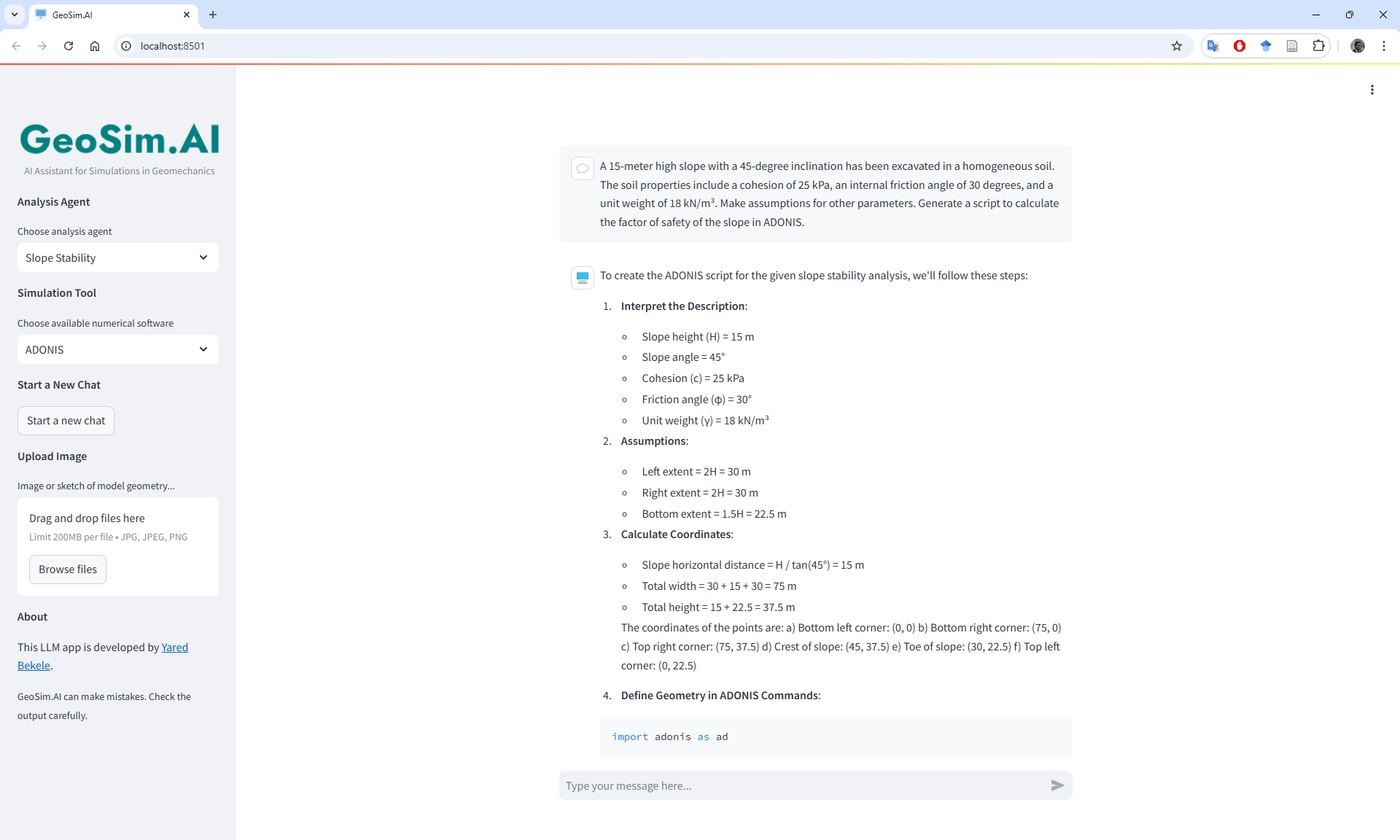}
    ~
    \includegraphics[scale=0.32]{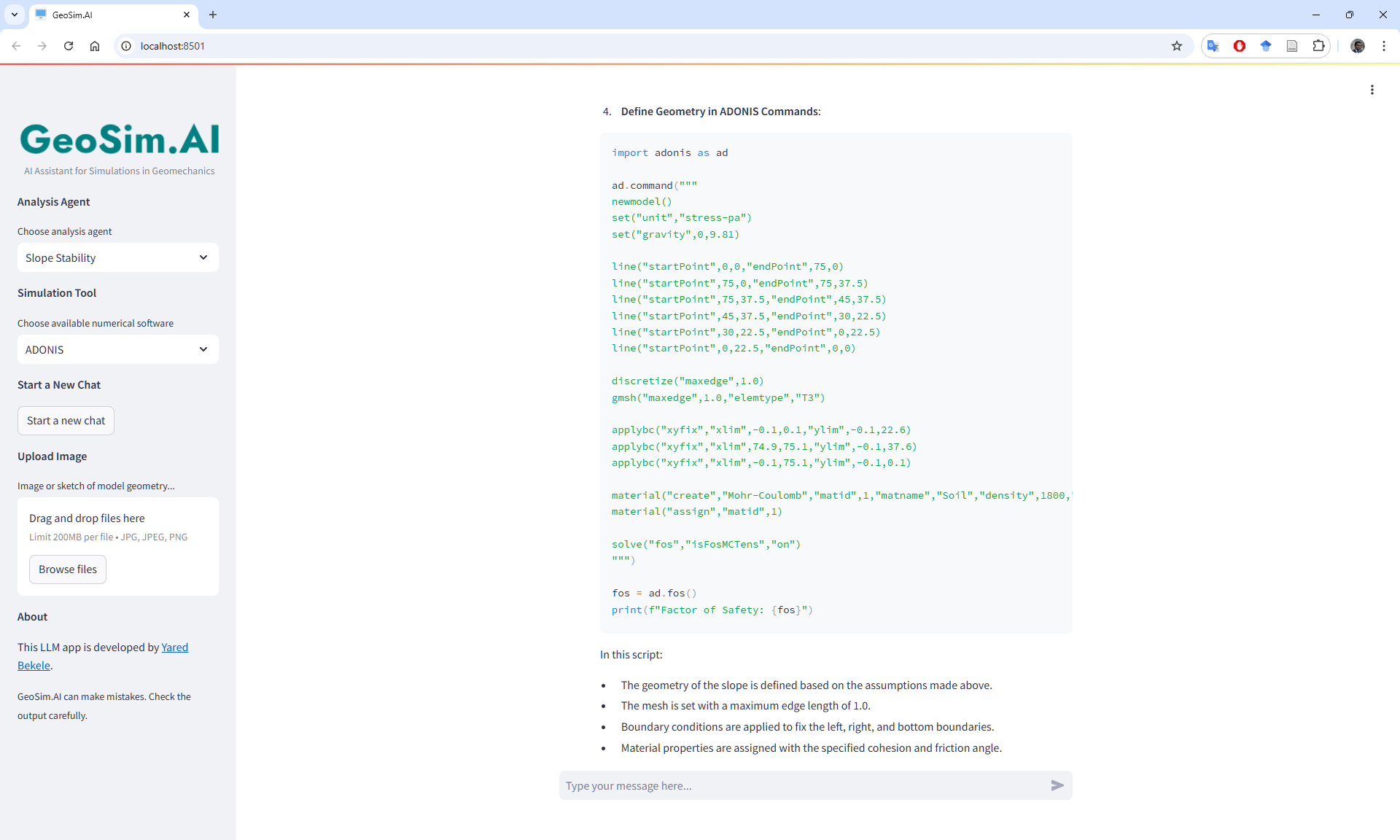}
    \caption{AI assistant for ADONIS - demo with text prompt.}
    \label{fig:ADONIS_text_prompt_demo}
\end{figure}

Our first demonstration focuses on a slope stability problem described in natural language as a text prompt. The response of the AI assistant is shown in Figure~\ref{fig:ADONIS_text_prompt_demo}. A video version of the demonstration can be viewed here: \href{https://youtu.be/Te3kfmKfaSA}{GeoSim.AI - ADONIS Agent Demo - Text Prompt}.

\subsubsection*{Image and Text Prompt}

\begin{figure}
    \centering
    \includegraphics[scale=0.32]{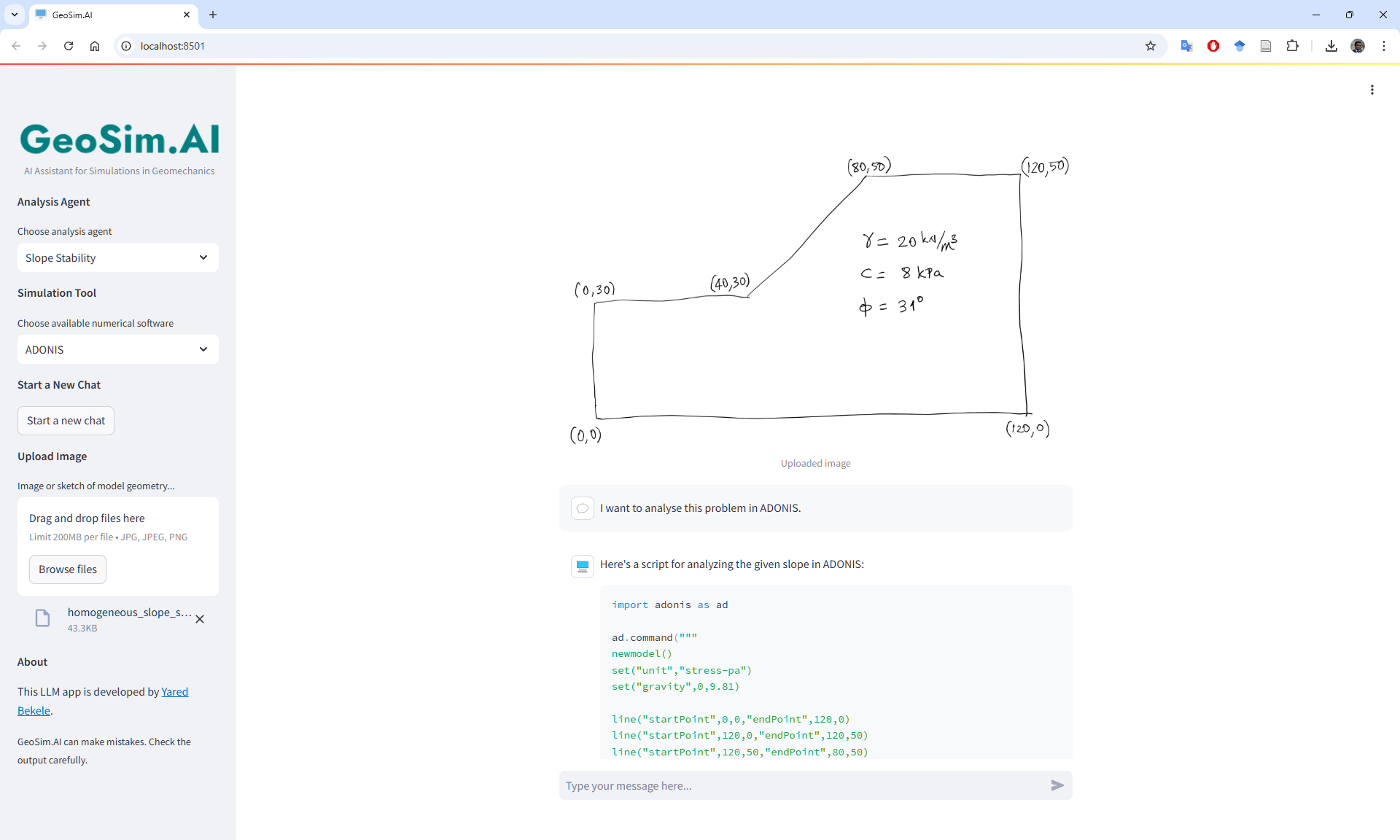}
    \includegraphics[scale=0.32]{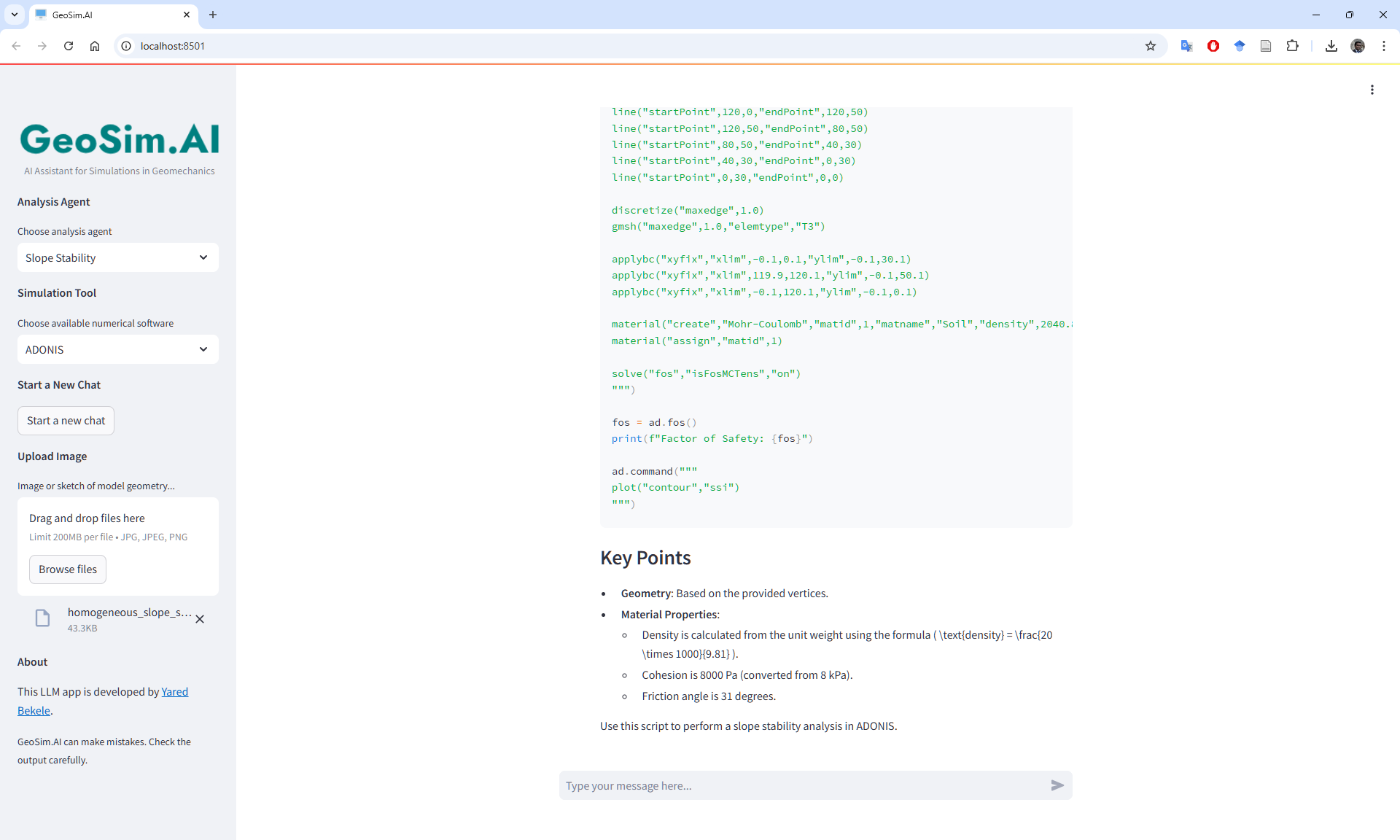}
    \caption{AI assistant for ADONIS - demo with image and text prompt.}
    \label{fig:ADONIS_image_and_text_prompt_demo}
\end{figure}

The next demo for the ADONIS AI assistant involves an image and text prompt where the slope stability problem sketch and material parameters are provided to GeoSim.AI via an image and instructed via a text instruction. The screenshots for this demo are shown in Figure~\ref{fig:ADONIS_image_and_text_prompt_demo}. A video version of the demonstration can be viewed here: \href{https://youtu.be/Uu2_jwBv4iw}{GeoSim.AI - ADONIS Agent Demo - Image + Text Prompt}.

\subsection{AI Assistant for HYRCAN}

\subsubsection*{Text Prompt}

\begin{figure}
    \centering
    \includegraphics[scale=0.32]{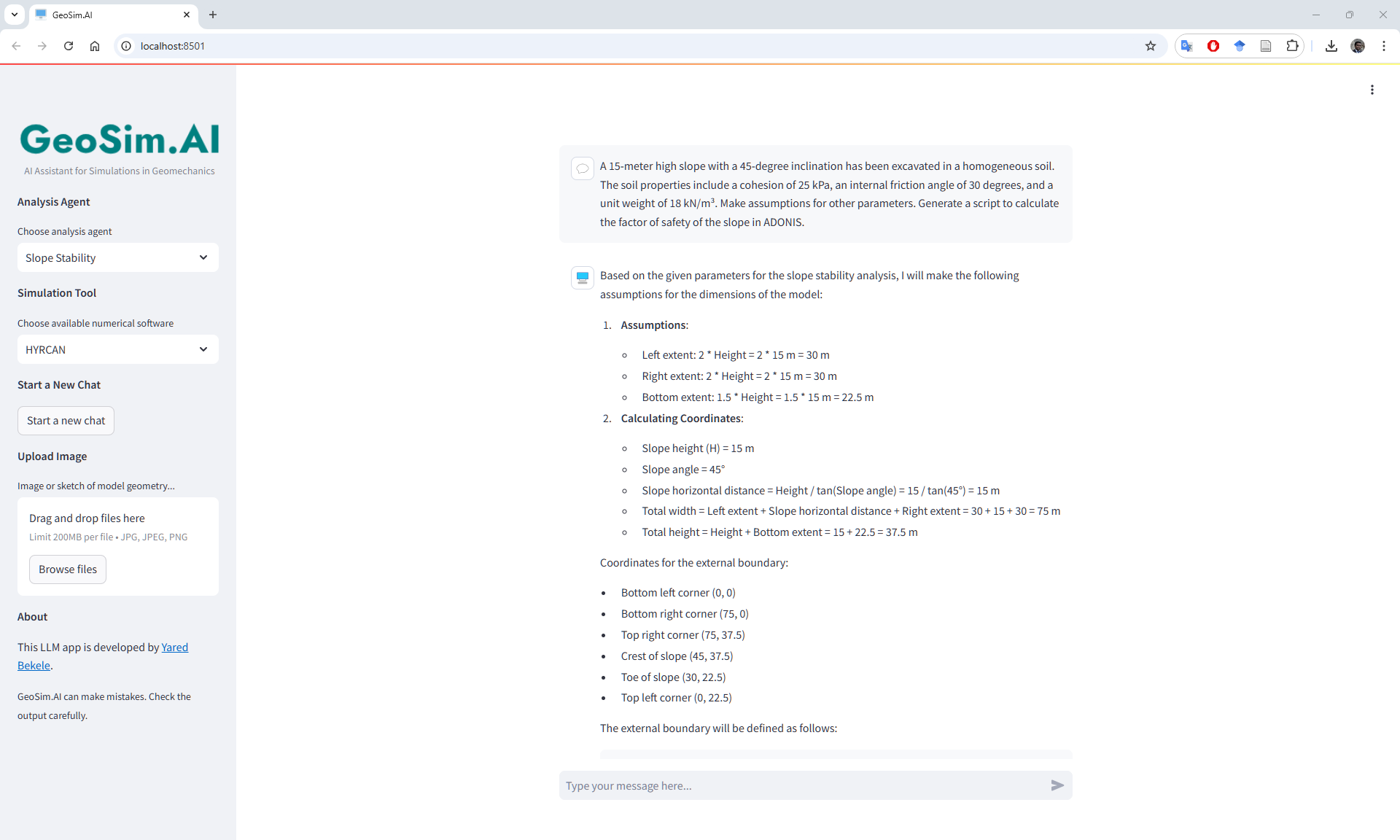}
    \includegraphics[scale=0.32]{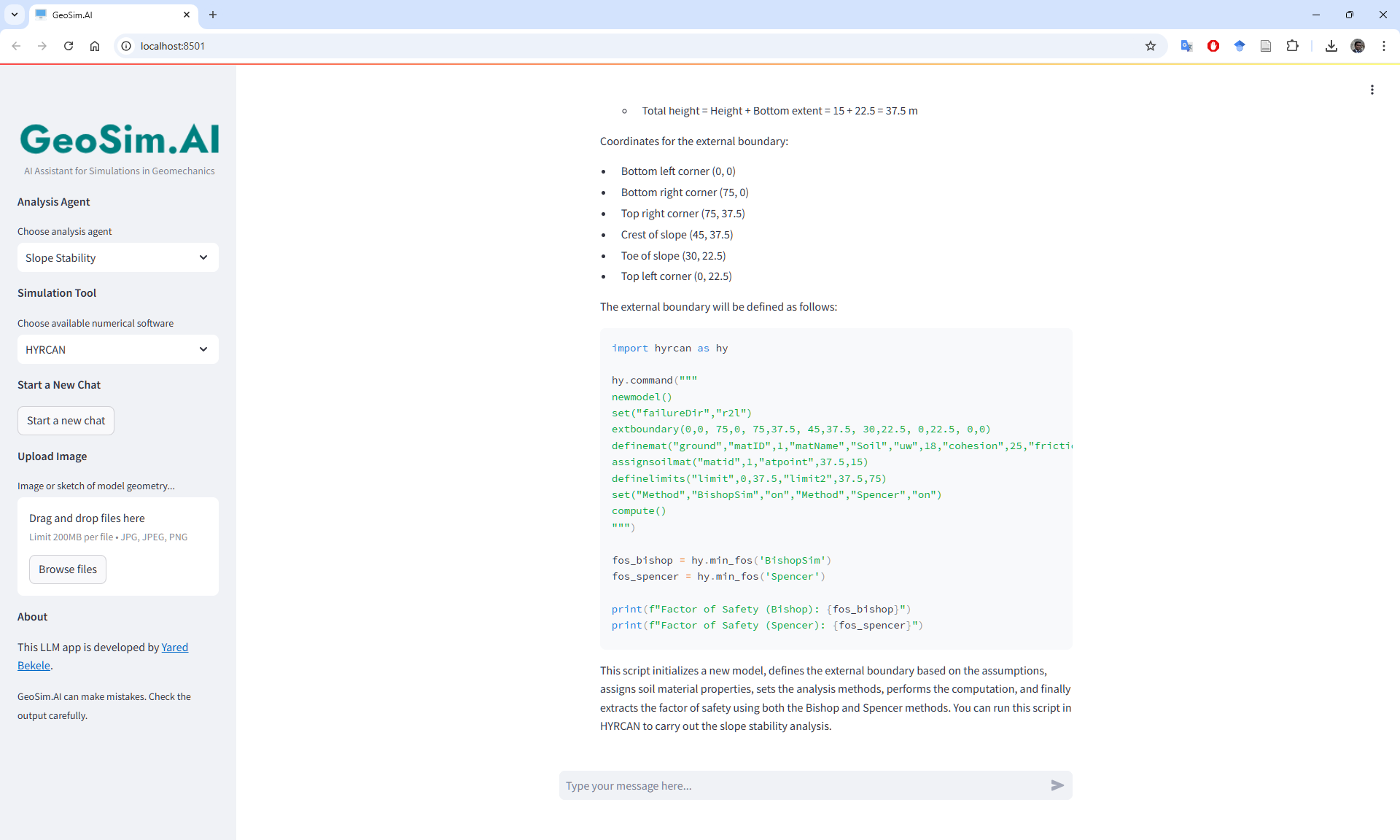}
    \caption{AI assistant for HYRCAN - demo with text prompt.}
    \label{fig:HYRCAN_text_prompt_demo}
\end{figure}

Demo with text prompt for the HYRCAN AI assistant is shown in Figure~\ref{fig:HYRCAN_text_prompt_demo}. A video version of the demonstration can be viewed by using the following link: \href{https://youtu.be/NVIivwbvIMg}{GeoSim.AI - HYRCAN Agent Demo - Text Prompt}.

\subsubsection*{Image and Text Prompt}

\begin{figure}
    \centering
    \includegraphics[scale=0.32]{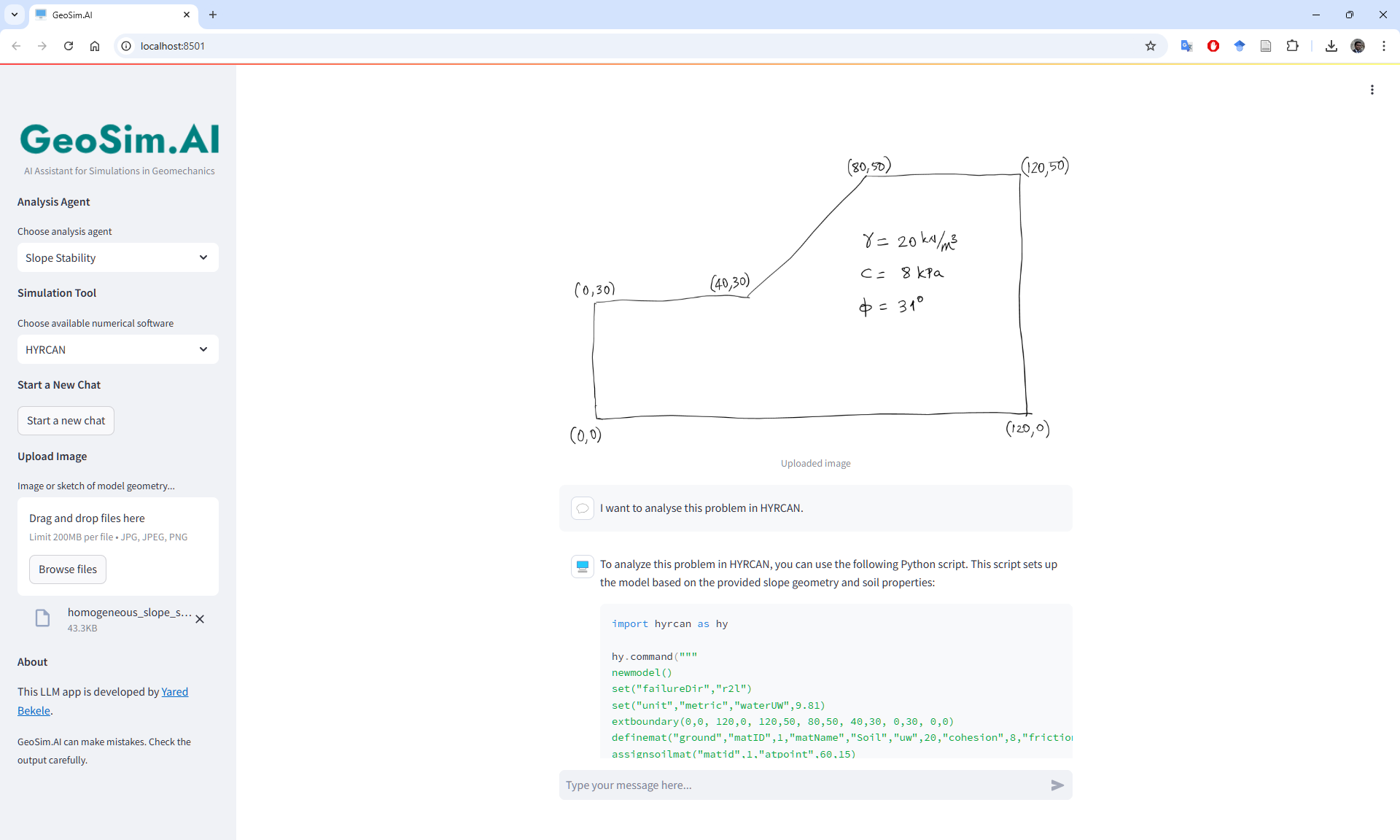}
    \includegraphics[scale=0.32]{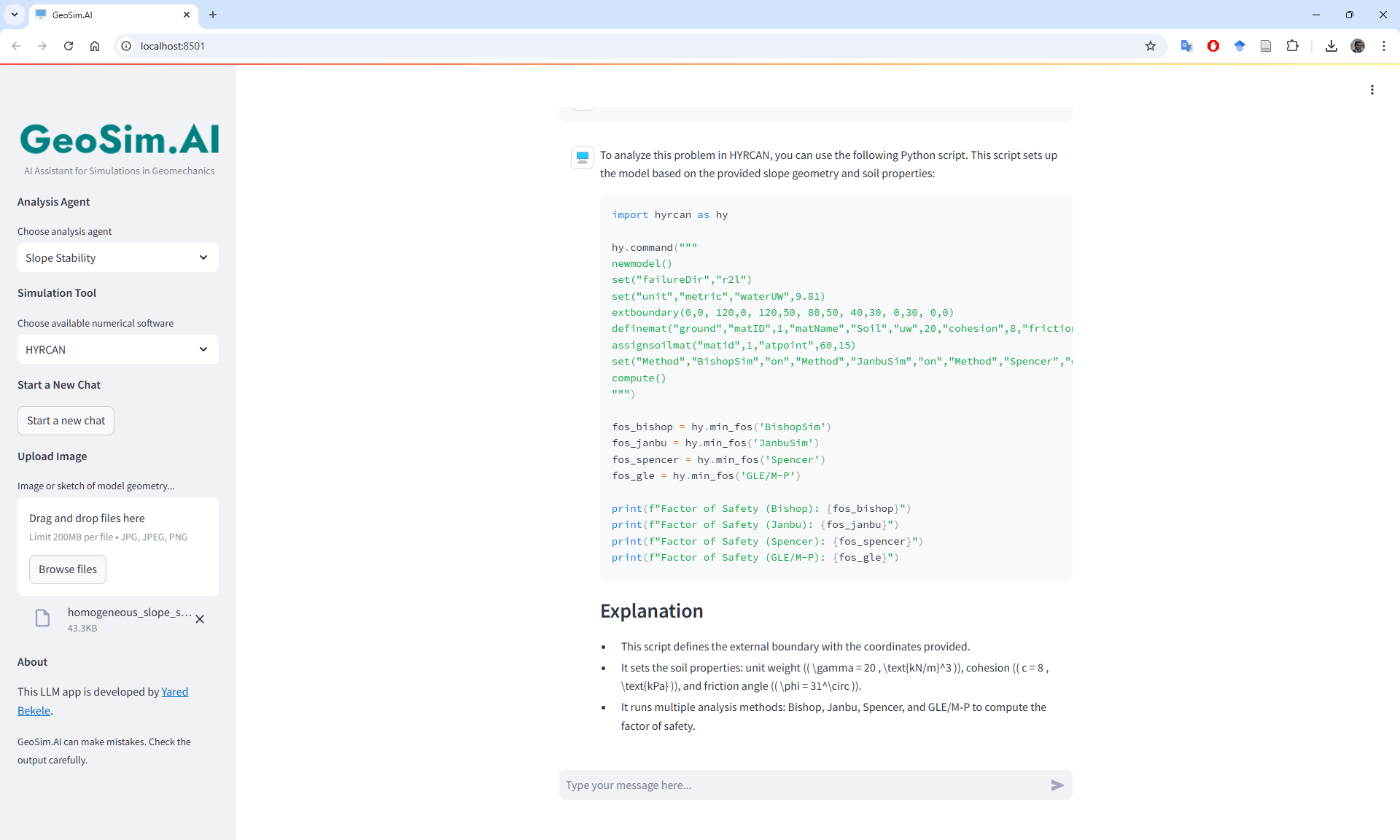}
    \caption{AI assistant for HYRCAN - demo with image and text prompt.}
    \label{fig:HYRCAN_image_and_text_prompt_demo}
\end{figure}

The screenshots for the demo with image and text prompt corresponding to the HYRCAN assitant are shown in Figure~\ref{fig:HYRCAN_image_and_text_prompt_demo}. A video version of the demonstration can be viewed by using the following link: \href{https://youtu.be/GPeWuJPa5Eg}{GeoSim.AI - HYRCAN Agent Demo - Image + Text Prompt}.

\subsection{AI Assistants for Commercial Software}

AI assistants for commercial software like FLAC2D, FLAC3D, PLAXIS2D and PLAXIS3D are also under development. Currently, the demonstrations and applications of these AI assistants are restricted to internal use only.

\FloatBarrier

\section{Cross-cutting Issues}

\subsection{Black Box within a Black Box?}

Computational software in geomechanics is often perceived as a "black box," particularly by users who lack a thorough understanding of its numerical algorithms and material models. The introduction of AI-driven numerical simulation assistants adds a new layer of abstraction, effectively creating what would seem like a black box within a black box. This development presents both challenges and opportunities. On one hand, it can make computational modeling more accessible to non-experts by automating complex processes and reducing the technical barriers to entry. On the other hand, it has the potential to significantly enhance productivity for experienced numerical modeling professionals by streamlining repetitive tasks and enabling them to focus on high-value aspects of their work. Such tools are most beneficial when used by experts with a strong theoretical understanding of numerical modeling principles and the ability to critically evaluate model assumptions, computational workflows, and results. By leveraging these technologies thoughtfully, experts can achieve greater efficiency while ensuring the accuracy and reliability of their simulations.

\subsection{Will Systems like GeoSim.AI Erode Expertise in Computational Geomechanics?}

The rise of AI-powered tools, such as GeoSim.AI as a suite of domain-specific AI assistants for numerical simulations in geomechanics, raises important questions about their impact on learning and understanding complex computational concepts. While these tools promise to improve productivity and reduce manual effort, they may also unintentionally encourage laziness among professionals, especially those who are less experienced or not motivated to deepen their knowledge of the field's core theories and methods.

A real concern is whether the ease provided by these AI systems could result in a generation of geomechanics professionals who are increasingly disconnected from the detailed mathematical and numerical ideas that form the foundation of their work. Those who are less engaged may become too dependent on these tools, treating them as black boxes instead of useful aids. However, this does not—and should not—apply to serious professionals.

Instead of lowering the value of expertise, these advancements should be seen as a chance to rethink how knowledge is applied. Skilled professionals can use AI tools to focus on solving bigger problems, creating new ideas, and tackling challenges that span multiple disciplines. For example, just as advanced calculators simplified routine arithmetic without eliminating the need for mathematical reasoning, or how Computer-Aided Design (CAD) transformed drafting while still requiring a strong understanding of design principles, AI tools in geomechanics can improve efficiency while keeping foundational knowledge vital.

In the end, it is up to each individual professional to use these tools as an opportunity to learn more rather than avoid learning. This requires a commitment to staying curious and actively engaging with the core ideas behind the work.

\section{Summary, Conclusions and Outlook}

This paper has introduced GeoSim.AI, a suite of AI assistants designed for numerical simulations in geomechanics. The system architecture of GeoSim.AI integrates a Large Language Model (LLM) core with two critical components: a comprehensive Knowledge Base containing geomechanical theory and practical information, and a Data \& Tools repository housing relevant datasets and simulation software tools. Through this integration, GeoSim.AI bridges the gap between natural language instructions and complex numerical simulations. The Knowledge Base, which forms the foundation of GeoSim.AI's technical understanding, is being continuously expanded through careful curation of geotechnical engineering knowledge and practical examples. The link between the LLM core and the Data \& Tools repository is currently under testing.

GeoSim.AI's implementation features a straightforward chat interface that accepts both text and image inputs. The interface allows users to select specific analysis agents and target software, making it adaptable to different simulation needs. The current version focuses primarily on slope stability problems, with successful demonstrations using both ADONIS and HYRCAN, which are open source geotechnical software packages. These demonstrations showed GeoSim.AI's capability to handle both pure text instructions and combinations of images and text to set up slope stability analyses effectively. Additionally, AI assistants for commercial software packages like FLAC2D, FLAC3D, PLAXIS2D and PLAXIS3D are under continuous development, with some existing result demonstrations and use cases that are currently restricted to internal use. 

GeoSim.AI and similar AI assistants raise two important issues that need careful consideration. First is the question of adding another layer of abstraction to already complex software. Computational software in geomechanics is often seen as a "black box," especially by users who don't fully understand its numerical methods and material models. While AI assistants can make computational modeling more accessible to non-experts by automating complex processes, they can also help experienced professionals work more efficiently. These tools work best when used by experts who understand numerical modeling principles and can properly evaluate the model assumptions, workflows, and results. The second issue concerns the impact on expertise in computational geomechanics. While these tools make work easier and faster, an issue of over-reliance and eroding of technical expertise may be raised, especially among less experienced users who might treat them simply as automated tools. However, this should be viewed similarly to how other technological advances have affected technical fields. Just as calculators didn't eliminate the need for mathematical understanding, or how Computer-Aided Design (CAD) still requires strong design principles, AI tools in geomechanics can improve efficiency while maintaining the importance of fundamental knowledge. The key is for professionals to use these tools to enhance their capabilities rather than replace their need to understand the underlying principles.

GeoSim.AI represents a significant advancement in how numerical simulations are performed in geomechanics. By translating natural language and image inputs into precise simulation commands, GeoSim.AI allows students, researchers, and engineers to focus on the physical aspects of geomechanical problems rather than spending time learning the specific rules and syntax of different software packages. This shift in focus from software technicalities to problem physics can lead to increased productivity and efficiency in computational geomechanics.

\section*{Acknowledgements}
The authors gratefully acknowledge the internal support from SINTEF, which was instrumental in the realization of this paper and the associated tool. The first author wishes to express his deep appreciation to his wife, Senait, for her unwavering support and patience during the development of this work, which required significant personal time, including numerous evenings and weekends.

\section*{Disclaimer}
All of the text in this paper is initially written by the author while general-purpose AI tools are later used for proofing and rephrasing without adding new text.

\bibliographystyle{unsrtnat}
\bibliography{geosimai}  






\end{document}